# COMMENTARY ON: "Citing and Reading Behavours in High-Energy Physics" (Gentil-Beccot et al 2009)

Stevan Harnad
Canada Research Chair in Cognitive Sciences
Université du Québec à Montréal
&
School of Electronics and Computer Science
University of Southampton

**ABSTRACT:** Evidence confirming that OA increases impact will not be sufficient to induce enough researchers to provide OA; only mandates from their institutions and funders can ensure that. HEP researchers continue to submit their papers to peer-reviewed journals, as they always did, depositing both their unrefereed preprints and their refereed postprints. None of that has changed. In fields like HEP and astrophysics, the journal affordability/accessibility problem is not as great as in many other fields, where it the HEP Early Access impact advantage translates into the OA impact advantage itself. Almost no one has ever argued that Gold OA provides a greater OA advantage than Green OA. The OA advantage is the OA advantage, whether Green or Gold.

Gentil-Beccot et al's (2009) study is an important one, and most of its conclusions are valid:

**(1)** Making research papers open access (OA) dramatically increases their impact.

**(2)** The earlier that papers are made OA, the greater their impact.

**(3)** High Energy Physics (HEP) researchers were among the first to make their papers OA (since 1991, and they did it without needing to be mandated to do it!)

> **(4)** Gold OA provides no further impact advantage over and above Green OA.

However, the following caveats need to be borne in mind, in interpreting this paper:

**(a)** HEP researchers have indeed been providing OA since 1991, unmandated (and computer scientists have been doing so since even earlier). But in the ensuing years, the only other discipline that has followed suit, unmandated, has been economics, despite the repeated demonstration of the Green OA impact advantage across all disciplines. So whereas still further evidence (as in this paper by Gentil-Beccot et al) confirming that OA increases impact is always very welcome, *that evidence will not be sufficient to induce enough researchers to provide OA*; only mandates from their institutions and funders can ensure that they do so.

**(b)** From the fact that *when there is a Green OA version available*, users prefer to consult that Green OA version rather than the journal version, it definitely does *not* follow that journals are no longer necessary. Journals are (and always were) essentially *peer-review service-providers and cerifiers*, and they still are. That essential function is indispensable. HEP researchers continue to submit their papers to peer-reviewed journals, as they always did; and they deposit both their unrefereed preprints and then their refereed postprints in arxiv (along with the journal reference). None of that has changed one bit.

**(c)** Although it has not been systematically demonstrated, it is likely that in fields like HEP and astrophysics, the journal affordability/accessibility problem is not as great as in many other fields. OA's most important function is to provide immediate access to those who cannot afford access to the journal version. Hence the Early Access impact advantage in HEP -- arising from making preprints OA well before the published version is available -- translates, in the case of most other fields, into the OA impact advantage itself, because *without OA many potential users simply do not have access even after publication, hence cannot make any contribution to the article's impact*.

**(d)** Almost no one has ever argued (let alone adduced evidence) that Gold OA provides a *greater* OA advantage than Green OA. *The OA advantage is the OA advantage*, whether Green or Gold. (It just happens to be easier and more rigorous to test and demonstrate the OA advantage through within-journal comparisons [i.e Green vs. non-Green articles] than between-journal comparisons [Gold vs. non-Gold journals].)